# A New Dual-Material Double-Gate (DMDG) Nanoscale SOI MOSFET – Two-dimensional Analytical Modeling and Simulation

G.Venkateshwar Reddy and M.Jagadesh Kumar[1]

Department of Electrical Engineering,

Indian Institute of Technology, Delhi,

Hauz Khas, New Delhi – 110 016, INDIA.

Email: mamidala@ieee.org   Fax: 91-11-2658 1264

---

[1]Author for Correspondence

**Abstract**

In this paper, we present the unique features exhibited by modified asymmetrical Double Gate (DG) silicon on insulator (SOI) MOSFET. The proposed structure is similar to that of the asymmetrical DG SOI MOSFET with the exception that the front gate consists of two materials. The resulting modified structure, Dual Material Double Gate (DMDG) SOI MOSFET, exhibits significantly reduced short channel effects when compared with the DG SOI MOSFET. Short channel effects in this structure have been studied by developing an analytical model. The model includes the calculation of the surface potential, electric field, threshold voltage and drain induced barrier lowering. A model for the drain current, transconductance, drain conductance and voltage gain is also discussed. It is seen that short channel effects in this structure are suppressed because of the perceivable step in the surface potential profile, which screens the drain potential. We further demonstrate that the proposed DMDG structure provides a simultaneous increase in the transconductance and a decrease in the drain conductance when compared with the DG structure. The results predicted by the model are compared with those obtained by two-dimensional simulation to verify the accuracy of the proposed analytical model.

# 1. Introduction

Double Gate (DG) MOSFETs using lightly doped ultra thin layers seem to be a very promising option for ultimate scaling of CMOS technology [1]. Excellent short-channel effect immunity, high transconductance and ideal subthreshold factor have been reported by many theoretical and experimental studies on this device [2-11]. In particular, asymmetrical DG SOI MOSFETs (front gate $p^+$ poly and back gate $n^+$ poly) are becoming popular since this type of structure provides a desirable threshold voltage (not too high or too low) unlike the symmetrical DG SOI MOSFETs.

The control of the gate voltage on the threshold voltage decreases as the channel length shrinks because of the increased charge sharing from source and drain. Therefore, the threshold voltage reduction with decreasing channel lengths and drain induced barrier lowering (DIBL) are important issues that need to be addressed while providing immunity against short-channel effects (SCEs) [5], [6]. To enhance the immunity against short channel effects, a new structure called dual material gate (DMG) MOSFET was proposed [12-15]. This structure has two metals in the gate M1 and M2 with different work functions. Such a configuration provides simultaneous increase in transconductance and suppressed short channel effects due to a step in the surface potential profile when compared with a single gate MOSFET. In the DMG structure, the peak electric field at the drain end is reduced, which ensures that the average electric field under the gate is increased. This enables an increased lifetime of the device, minimization of the ability of the localized charges to raise drain resistance [16] and more control of gate over the conductance of the channel so as to increase the gate transport efficiency. The step function profile of the surface potential ensures screening of the channel region under the material on the source side (M1) from drain potential variations. After saturation, M2 absorbs any additional drain-source voltage and hence the region under M1 is screened from drain potential variations.

However, the drivability, and the transconductance of the DMG structure are not as good as that of the DG structure.

To incorporate the advantages of both DG and DMG structures, we propose a new structure, Dual-Material Double-Gate (DMDG) SOI MOSFET that is similar to that of an asymmetrical DG SOI MOSFET with the exception that the front gate of the DMDG structure consists of two materials ($p^+$ poly and $n^+$ poly). The aim of this paper is, therefore, to present using two-dimensional simulation, the reduced short channel effects exhibited by the DMDG structure below 100nm, while simultaneously achieving a higher transconductance and reduced drain conductance compared to the DG SOI MOSFET. The proposed structure exhibits the desired features of both the DMG and the DG structures. With this structure, we demonstrate a considerable reduction in the peak electric field near the drain end, increased drain breakdown voltage, improved transconductance, reduced drain conductance and a desirable threshold voltage "roll-up" even for channel lengths far below 100 nm. An analytical model using Poisson's equation also has been presented for the surface potential leading to the threshold voltage model for the DMDG SOI MOSFET. A complete drain current model [18] considering impact ionization [19], velocity overshoot, channel length modulation and DIBL [20] is also presented. The accuracy of the model is verified by comparing the model results with the simulation results using a 2-D device simulator, MEDICI[17].

## 2. Analytical Model for Surface Potential

Schematic cross-sectional views of both asymmetrical DG and DMDG SOI MOSFET implemented using the 2-D device simulator MEDICI is shown in Fig. 1. The front gate consists of dual materials M1 ($p^+$ poly) and M2 ($n^+$ poly) of lengths $L_1$ and $L_2$ respectively, while the back gate is effectively an $n^+$ poly gate. Assuming the impurity density in the channel region to be uniform, and neglecting the effect of the fixed oxide charges on the electrostatics of the channel, the potential distribution in the silicon thin film before the onset of strong inversion can be written as [21]:

$$\frac{d^2\phi(x,y)}{dx^2} + \frac{d^2\phi(x,y)}{dy^2} = \frac{qN_A}{\varepsilon_{si}} \quad \text{for } 0 \leq x \leq L,\ 0 \leq y \leq t_{si} \quad (1)$$

where $N_A$ is the uniform film doping concentration independent of the gate length, $\varepsilon_{si}$ is the dielectric constant of silicon, $t_{si}$ is the film thickness and $L$ is the device channel length. The potential profile in the vertical direction, i.e., the y-dependence of $\phi(x,y)$ can be approximated by a simple parabolic function as proposed by [21] for fully depleted SOI MOSFET's as

$$\phi(x,y) = \phi_s(x) + a_1(x)y + a_2(x)y^2 \quad (2)$$

where $\phi_s(x)$ is the surface potential and the arbitrary coefficients $a_1(x)$ and $a_2(x)$ are functions of $x$ only.

In a DG-SOI MOSFET the front gate consists of only one material i.e, $p^+$ poly, but in the DMDG structure, we have two different materials ($p^+$ poly and $n^+$ poly) with work functions $\phi_{M1}$ and $\phi_{M2}$, respectively. Therefore, the front channel flat-band voltages of the $p^+$ poly and $n^+$ poly at the front gate would be different and they are given as

$$V_{FB,fp} = \phi_{MS1} = \phi_{M1} - \phi_{Si} \quad \text{and} \quad V_{FB,fn} = \phi_{MS2} = \phi_{M2} - \phi_{Si} \quad (3)$$

where $\phi_{si}$ is the silicon work function which is given by

$$\phi_{Si} = \chi_{Si} + \frac{E_g}{2q} + \phi_F \quad (4)$$

where $E_g$ is the silicon bandgap at 300K, $\chi_{si}$ is the electron affinity of silicon, $\phi_F = V_T \ln(N_A/n_i)$ is the Fermi potential, $V_T$ is the thermal voltage and $n_i$ is the intrinsic carrier concentration. Since we have two regions in the front gate of the DMDG structure, the surface potential under $p^+$ poly and $n^+$ poly can be written based on (2) as:

$$\phi_1(x, y) = \phi_{s1}(x) + a_{11}(x)y + a_{12}(x)y^2 \quad \text{for} \quad 0 \leq x \leq L_1, \quad 0 \leq y \leq t_{si} \quad (5)$$

$$\phi_2(x, y) = \phi_{s2}(x) + a_{21}(x)y + a_{22}(x)y^2 \quad \text{for} \quad L_1 \leq x \leq L_1 + L_2, \quad 0 \leq y \leq t_{si} \quad (6)$$

where $\phi_{s1}$ and $\phi_{s2}$ are the surface potentials under $p^+$ poly (M1) and $n^+$ poly (M2) respectively and $a_{11}$, $a_{12}$, $a_{21}$ and $a_{22}$ are arbitrary coefficients.

The Poisson's equation is solved separately under the two top front gate materials ($p^+$ poly and $n^+$ poly) using the following boundary conditions:

1. Electric flux at the front gate-oxide interface is continuous for the dual material gate. Therefore, we have

$$\left.\frac{d\phi_1(x, y)}{dy}\right|_{y=0} = \frac{\varepsilon_{ox}}{\varepsilon_{si}} \frac{\phi_{s1}(x) - V'_{GS,f1}}{t_f} \quad \text{under M1} \quad (7)$$

$$\left.\frac{d\phi_2(x, y)}{dy}\right|_{y=0} = \frac{\varepsilon_{ox}}{\varepsilon_{si}} \frac{\phi_{s2}(x) - V'_{GS,f2}}{t_f} \quad \text{under M2} \quad (8)$$

where $\varepsilon_{ox}$ is the dielectric constant of the oxide, $t_f$ is the gate oxide thickness and

$$V'_{GS,f1} = V_{GS} - V_{FB,fp} \quad \text{and} \quad V'_{GS,f2} = V_{GS} - V_{FB,fn} \quad (9)$$

where $V_{GS}$ is the gate-to-source bias voltage, $V_{FB,fp}$ and $V_{FB,fn}$ are the front-channel flat-band voltages of $p^+$ polysilicon and $n^+$ polysilicon, respectively, and are given by (3).

2. Electric flux at the back gate-oxide and the back channel interface is continuous for both the materials of the front gate ($p^+$ poly and $n^+$ poly).

$$\left.\frac{d\phi_1(x, y)}{dy}\right|_{y=t_{si}} = \frac{\varepsilon_{ox}}{\varepsilon_{si}} \frac{V'_{GS,b} - \phi_B(x)}{t_b} \quad \text{under M1} \quad (10)$$

$$\left.\frac{d\phi_2(x, y)}{dy}\right|_{y=t_{si}} = \frac{\varepsilon_{ox}}{\varepsilon_{si}} \frac{V'_{GS,b} - \phi_B(x)}{t_b} \quad \text{under M2} \quad (11)$$

where $t_b$ is the back gate oxide thickness, $\phi_B(x)$ is the potential function along the back gate oxide-silicon interface, and $V'_{GS,b} = V_{GS} - V_{FB,bn}$, where $V_{FB,bn}$ is the back gate flat-band voltage and is same as that of $V_{FB,fn}$.

3. Surface potential at the interface of the two dissimilar gate materials of the front gate is continuous

$$\phi_1(L_1, 0) = \phi_2(L_1, 0) \tag{12}$$

4. Electric flux at the interface of two materials of the front gate is continuous

$$\left.\frac{d\phi_1(x,y)}{dx}\right|_{x=L_1} = \left.\frac{d\phi_2(x,y)}{dx}\right|_{x=L_1} \tag{13}$$

5. The potential at the source end is

$$\phi_1(0,0) = \phi_{s1}(0) = V_{bi} \tag{14}$$

where $V_{bi} = V_T \ln\left(\frac{N_A N_D}{n_i^2}\right)$ is the built-in potential across the body-source junction and $N_A$ and $N_D$ are the body and source/drain dopings respectively.

6. The potential at the drain end is

$$\phi_2(L_1 + L_2, 0) = \phi_{s2}(L_1 + L_2) = V_{bi} + V_{DS} \tag{15}$$

where $V_{DS}$ is the applied drain-source bias. The constants $a_{11}(x)$, $a_{12}(x)$, $a_{21}(x)$ and $a_{22}(x)$ in (5) and (6) can be found from the boundary conditions (7)-(11). Substituting these constants in (5) and (6) and then in (1) we get

$$\frac{d^2\phi_{s1}(x)}{dx^2} + \alpha\phi_{s1}(x) = \beta_1 \quad \text{and} \quad \frac{d^2\phi_{s2}(x)}{dx^2} + \alpha\phi_{s2}(x) = \beta_2 \tag{16}$$

where
$$\alpha = \frac{2(1 + C_f/C_b + C_f/C_{si})}{t_{si}^2(1 + 2C_{si}/C_b)}$$

$$\beta_1 = \frac{qN_A}{\varepsilon_{si}} - 2V'_{GS,f1}\left(\frac{C_f/C_b + C_f/C_{si}}{t_{si}^2(1+2C_{si}/C_b)}\right) - 2V'_{GS,b}\left(\frac{1}{t_{si}^2(1+2C_{si}/C_b)}\right)$$

$$\beta_2 = \frac{qN_A}{\varepsilon_{si}} - 2V'_{GS,f2}\left(\frac{C_f/C_b + C_f/C_{si}}{t_{si}^2(1+2C_{si}/C_b)}\right) - 2V'_{GS,b}\left(\frac{1}{t_{si}^2(1+2C_{si}/C_b)}\right)$$

where $C_{si} = \varepsilon_{si}/t_{si}$, $C_f = \varepsilon_{ox}/t_f$ and $C_b = \varepsilon_{ox}/t_b$.

The above equations are second order differential equations with constant coefficients and the expression for surface potential under p$^+$ poly and n$^+$ poly of the front gate is of the form

$$\phi_{s1}(x) = A\exp(\eta x) + B\exp(-\eta x) - \beta_1/\alpha \qquad \text{for} \qquad 0 \leq x \leq L_1 \qquad \text{under M1} \qquad (17)$$

$$\phi_{s2}(x) = C\exp(\eta(x-L_1)) + D\exp(-\eta(x-L_1)) - \beta_2/\alpha \qquad \text{for} \quad L_1 \leq x \leq L \quad \text{under M2} \quad (18)$$

where $\eta = \sqrt{\alpha}$ and $L=L_1+L_2$. Using boundary conditions (12)-(15), we obtain A, B, C, and D as

$$A = \left((V_{bi} - \sigma_2 + V_{DS}) - (V_{bi} - \sigma_1)\exp(-\eta L) - (\sigma_1 - \sigma_2)\cosh(\eta L_2)\right)\left\{\frac{\exp(-\eta L)}{1-\exp(-2\eta L)}\right\}$$

$$B = \left((V_{bi} - \sigma_1) - (V_{bi} - \sigma_2 + V_{DS})\exp(-\eta L) + (\sigma_1 - \sigma_2)\cosh(\eta L_2)\right)\left\{\frac{\exp(-\eta L)}{1-\exp(-2\eta L)}\right\}$$

$$C = A\exp(\eta L_1) + \frac{(\sigma_1 - \sigma_2)}{2} \qquad \text{and} \qquad D = B\exp(-\eta L_1) + \frac{(\sigma_1 - \sigma_2)}{2}$$

where $\sigma_1 = -\beta_1/\alpha$ and $\sigma_2 = -\beta_2/\alpha$.

The electric field distribution along the channel length can be obtained by differentiating the surface potential given by (17) and (18) and can be written as:

$$E_1(x) = \left.\frac{d\phi_1(x,y)}{dx}\right|_{y=0} = A\eta\exp(\eta x) - B\eta\exp(-\eta x) \qquad 0 \leq x \leq L_1 \qquad \text{under M1} \qquad (19)$$

$$E_2(x) = \left.\frac{d\phi_2(x,y)}{dx}\right|_{y=0} = C\eta\exp(\eta(x-L_1)) - D\eta\exp(-\eta(x-L_1)) \quad L_1 \leq x \leq L \quad \text{under M2} \quad (20)$$

The above two equations are quite useful in determining how the drain side electric field is modified by the proposed DMDG structure.

## 3. Threshold Voltage and DIBL model for the DMDG SOI MOSFET

In the proposed DMDG SOI MOSFET, we have $t_f$ and $t_b$ as the front gate and back gate oxide thicknesses, and the same gate voltage, $V_G$, is applied to both the gates. The channel doping is uniform with an acceptor concentration of $10^{15}$ cm$^{-3}$ as in [18]. The threshold voltage, $V_{th}$ for the DMDG SOI structure is derived from the graphical approach as has been done for DG SOI MOSFETs in [18]. When the potential distribution dependence on the gate voltage is studied, it is seen that first an inversion layer is formed on the inside surface of the back gate n$^+$ polysilicon. Then the potential distribution changes linearly while the surface potential is fixed. After this an inversion layer on the inside surface of the p$^+$ polysilicon is formed and then the potential distribution in the channel is invariable and the applied voltage is sustained by both the gate oxides. This analysis concludes that this structure has two different threshold voltages related to the front and the back gate respectively.

Based on the graphical approach from [18], the expression for the front and the back gate threshold voltage of the long channel device is given as

$$V_{th1} = V_{FB,fp} + 2\phi_F + \frac{Q_{si}}{2}\left(1 + V_T \frac{4C_{si}}{Q_{si}}\right)\left(\frac{1}{4C_{si}} + \frac{1}{C_f}\right) + V_T \ln\left(V_T \frac{4C_{si}}{Q_{si}}\right) - \frac{\gamma t_f + t_{si}}{\gamma t_f + \gamma t_b + t_{si}} \Delta V_{FB} \quad (21)$$

$$V_{th2} = V_{FB,fp} + 2\phi_F + \frac{Q_{si}}{2}\left(1 + V_T \frac{4C_{si}}{Q_{si}}\right)\left(\frac{1}{4C_{si}} + \frac{1}{C_f}\right) + V_T \ln\left(V_T \frac{4C_{si}}{Q_{si}}\right) \quad (22)$$

where $V_{th1}$ is the threshold voltage for the back gate with n$^+$ poly and $V_{th2}$ is the threshold voltage for the front gate with p$^+$ poly and n$^+$ poly, $V_{FB,fp}$ and $V_{FB,fn}$ are given by (3), $\gamma = \varepsilon_{si}/\varepsilon_{ox}$, $Q_{si} = qN_A t_{si}$ is the channel acceptor charge and $\Delta V_{FB}$ is the difference between flatband voltages associated with the front and the back gates and is given by

$$\Delta V_{FB} = V_{FB,fp} - V_{FB,bn} \quad (23)$$

In the above models, both the induced and the depleted charges have been considered in the channel region. However, for short channel devices, we neglect both the charges in the derivation of the threshold voltage model. Low doping concentration of the double-gate SOI MOSFETs makes this a good approximation [22]. This approximation leads to a Poisson equation of potential, ϕ, given by

$$\frac{d^2\phi(x,y)}{dx^2} + \frac{d^2\phi(x,y)}{dy^2} = \frac{qN_A}{\varepsilon_{si}} \approx 0 \tag{24}$$

As in [22], the above equation can be solved using the parabolic potential profile (5) and (6) and with the help of the boundary conditions (7)-(15). The short channel threshold voltage shift $\Delta V_{th}$ of the DMDG SOI MOSFET can be given as

$$\Delta V_{th} = 2\sqrt{\eta_S \eta_{L1}} e^{-\zeta} \tag{25}$$

where

$$\eta_S = V_{bi} - V_{GS,f1}' + \frac{\Delta V_{FB}}{2\left(1 + \frac{t_{si}}{2\gamma t_f}\right)} \tag{26}$$

$$\eta_{L1} = \frac{1}{2}\left[\frac{\left(V_{bi} + V_{DS} - V_{GS,f2}'\right)\sinh\left(\frac{L_1}{\lambda}\right) + \eta_S \sinh\left(\frac{L_2}{\lambda}\right)}{\cosh\left(\frac{L_1}{\lambda}\right)\sinh\left(\frac{L_2}{\lambda}\right) + \sinh\left(\frac{L_1}{\lambda}\right)\cosh\left(\frac{L_2}{\lambda}\right)}\right] \tag{27}$$

$$\zeta = L_1 / \sqrt{2\gamma t_{si} t_f} \tag{28}$$

Therefore, the expression for the threshold voltage of the DMDG SOI MOSFET is given by

$$V_{th} = V_{thL} - \Delta V_{th} \tag{29}$$

where $V_{thL}$ can be either $V_{th1}$ or $V_{th2}$. Equation (29) does not predict any threshold voltage roll-up because the coupling between the front dual material gate and the back gate is not considered in the above model. However, as we will demonstrate, based on simulation, in the following sections, the DMDG structure does exhibit a small threshold voltage roll-up phenomenon. To take this roll-up into account, we introduce an empirical correction factor, $\theta$. The empirical relation used for this parameter is given by

$$\theta = 1 - \frac{L_1 - L_2}{\rho L_1} \quad (30)$$

Here the value of $\rho$, when compared with the simulated results has been obtained as $\rho = kL_1 - 2.25$, where $L_1$ is in $\mu m$ and $k=185/\mu m$. Therefore, the final expression for the threshold voltage of the DMDG SOI MOSFET is given by

$$V_{th} = V_{thL} - \theta \Delta V_{th} \quad (31)$$

It is to be noted that when $L_1 = L_2$, $\theta$ is equal to unity and when the channel length is reduced keeping $L_1$ fixed, then $\theta$ decreases leading to a threshold voltage roll-up. It is assumed here that the length of M1 is always greater than that of M2, which is reasonable for sub 100 nm channel lengths.

Using the threshold model given by (31), the DIBL of the DMDG structure can now be expressed as

$$DIBL = V_{th}(V_{DS} = 0) - V_{th}(V_{DS}) = V_{th,lin} - V_{th,sat} \quad (32)$$

where $V_{th,lin}$ and $V_{th,sat}$ are the threshold voltages in the linear and the saturation regimes, respectively.

## 4. IV Model

In order to derive the current-voltage characteristics, the proposed DMDG structure can be treated as two transistors connected in parallel, each having its own threshold voltage: $V_{th1}$ and $V_{th2}$ relating to the back gate and the front gate, respectively. The channel current is then given

by [18]

$$I_{ch} = \sum_{i=1,2} \frac{W\mu_{neff}C_{ox}}{L\left(1+\frac{V_{DS}}{LE_C}\right)}\left[(V_{GS}-V_{thi})V_{DS} - \frac{1}{2}V_{DS}^2\right] \quad \text{in linear region} \quad (33)$$

$$I_{ch} = \sum_{i=1,2} \frac{W\mu_{neff}C_{ox}}{L\left(1+\frac{V_{DS,sati}}{LE_C}\right)}\left[(V_{GS}-V_{thi})V_{DS,sati} - \frac{1}{2}V_{DS,sati}^2\right] \quad \text{in saturation region} \quad (34)$$

where $V_{thi}$ corresponds to the back gate and the front gate threshold voltage for $i = 1$ and 2, respectively. $E_C$ is the critical electric field at which the electron velocity ($\upsilon_{ns}$) saturates and $V_{DS,sati}$ is the saturation voltage and are given by

$$E_C = \frac{2\upsilon_{ns}}{\mu_{neff}} \; ; \qquad V_{DS,sati} = \frac{V_{GS}-V_{thi}}{1+\frac{V_{GS}-V_{thi}}{LE_C}} \quad (35)$$

where $\mu_{neff}$ is the effective mobility of the inversion layer electrons given by

$$\frac{1}{\mu_{neff}} = \frac{1}{\mu_{ph}} + \frac{1}{\mu_{sr}} \quad (36)$$

where $\mu_{ph}$ is the mobility associated with the phonon scattering and $\mu_{sr}$ is the mobility associated with the surface roughness scattering as discussed in [18]. However, (33) and (34) do not include the short channel effects, the parasitic BJT effects and the impact ionization. To develop an accurate analytical drain current model, we need to consider the above effects as discussed below.

*3.1 Impact ionization and parasitic BJT effects*

As the lateral electric field in the device is large in the saturation region, the impact ionization and the parasitic BJT effects strongly affect the current conduction of the device. In the inversion layer, at the oxide-silicon interface, there is a channel current ($I_{ch}$), which is due to the drifting of the electrons. In the high electric field region near the drain, the drifting electrons collide with the lattice resulting in the generation of electron-hole pairs. Because of the electric field, the electrons move towards the drain contact and the holes move in the source direction resulting in the impact ionization current ($I_h$). For a very short-channel SOI MOS device, the parasitic BJT with its emitter at the source and its collector at the drain cannot be overlooked. A portion of the impact ionization current ($KI_h$) is directed towards the source. As a result holes get accumulated in the thin film, which leads to the activation of the parasitic bipolar transistor. As the bipolar device is activated, these holes recombine with electrons in the base region. In the parasitic bipolar device, a portion of the collector current ($K'I_C$), which is mainly composed of electrons, is a result of the vertical electric field. These electrons also collide with lattice, and consequently generate electron-hole pairs.

Therefore, the drain current ($I_D$), considering the impact ionization and the parasitic BJT effects, has the following components: the channel current ($I_{ch}$), the impact ionization current ($I_h$), and the collector current ($I_C$) of the parasitic bipolar device [19]:

$$I_D = I_{ch} + I_h + I_C \qquad (37)$$

Substituting for $I_h$ and $I_C$ from [19], the expression for the drain current $I_D$ in saturation is given by

$$I_{D,sat} = GI_{ch} + HI_{CBO} \qquad (38)$$

where 
$$G = 1 + \frac{(M-1)\left[1-(1-K)\alpha_0\right]}{1-\left[1+KK'(M-1)\right]\alpha_0} \qquad H = \frac{1+K'(M-1)}{1-\left[1+KK'(M-1)\right]\alpha_0}$$

where $M$, $K$, $K'$, $\alpha_0$ and $I_{CBO}$ are as given in [19]. However, before the onset of saturation, the drain current is equal to the channel current, $I_{ch}$ given by (33).

*3.2 The channel length modulation, velocity overshoot and DIBL effects*

Non-local effects such as channel length modulation, velocity overshoot and DIBL are becoming more prominent as MOSFET dimensions shrink to the deep submicrometer regime and it is necessary to include them in the drain current model. Velocity overshoot is one of the most important effects from the practical point of view since it is directly related to the increase of current drive and transconductance as experimentally observed in short channel MOSFETs [23-26]. It has been shown that an electric field step causes the electron velocity to overshoot its saturation value for a period shorter than the energy relaxation time. Therefore, as the longitudinal electric field increases, the electron gas starts to be in non-equilibrium with the lattice with the result that electrons can be accelerated to velocities higher than the saturation velocity for channel lengths under 0.15 μm.

Using (33), (34) and considering velocity overshoot effects [27], the channel length modulation [28], and the DIBL [20], final expression for the channel current of the DMDG structure is given by

$$I_{ch} = \sum_{i=1,2} \left( \frac{W\mu_{neff}C_{ox}}{L\left(1-\frac{l_d}{L}+\frac{V_{DS}}{LE_C}\right)} \left[ \left(V_{GS}-V'_{thi}\right)V_{DS} - \frac{1}{2}V_{DS}^2 \right] + \lambda_a \frac{W}{(L-l_d)^2} \left[ \left(V_{GS}-V'_{thi}\right)V_{DS} - \frac{1}{2}V_{DS}^2 \right] \right) \quad (39)$$

$$I_{ch,sat} = \sum_{i=1,2} \left( \frac{W\mu_{neff}C_{ox}}{L\left(1-\frac{l_d}{L}+\frac{V_{DS,sati}}{LE_C}\right)} \left[ \left(V_{GS}-V'_{thi}\right)V_{DS,sati} - \frac{1}{2}V_{DS,sati}^2 \right] + \lambda_a \frac{W}{(L-l_d)^2} \left[ \left(V_{GS}-V'_{thi}\right)V_{DS,sati} - \frac{1}{2}V_{DS,sati}^2 \right] \right)$$
(40)

where (39) corresponds to the current in the linear region and (40) corresponds to the current in the saturation region, $\lambda_a$ is a parameter that takes into account velocity overshoot effects which is taken to be as $25\times10^{-5}$ cm$^3$/Vs as suggested in [27], $V'_{thi} = V_{thi} - DIBL$ and $l_d$ is the channel length modulation factor as given in [28].

*3.3 Total drain current*

Using (38) and (40), the total drain current of the DMDG SOI MOSFET is given by the expression

$$I_{D,sat} = G \sum_{i=1,2} \left( \frac{W \mu_{neff} C_{ox}}{L\left(1 - \frac{l_d}{L} + \frac{V_{DS,sati}}{LE_C}\right)} \left[ (V_{GS} - V'_{thi}) V_{DS,sati} - \frac{1}{2} V^2_{DS,sati} \right] + \lambda_a \frac{W}{(L-l_d)^2} \left[ (V_{GS} - V'_{thi}) V_{DS,sati} - \frac{1}{2} V^2_{DS,sati} \right] \right) + HI_{CBO}$$

(41)

Equation (41) corresponds to the drain current in the saturation region. Drain current in the linear region is equal to the channel current given by (39).

## 5. Results and Discussion

The two-dimensional device simulator MEDICI [17] was used to verify the proposed model for the DMDG structure. Typical dimensions used for both the DMDG and the DG structures are summarized in Table. 1. The surface potential distribution within the silicon thin-film was simulated with MEDICI. Fig. 2 shows the calculated and the simulated surface potential profile for a channel length of 100 nm ($L_1 = L_2 = 50$ nm) at the silicon-oxide interface of the DMDG structure along with the simulated potential profile of the DG structure. It is clearly seen that the DMDG structure exhibits a step function in the surface potential along the channel. Because of this unique feature, the area under $p^+$ poly front gate of the DMDG structure is essentially screened from the drain potential variations. This means that the drain potential has very little effect on the drain current after saturation [29] reducing the drain conductance and the drain induced barrier lowering (DIBL) as discussed below. The predicted values of the model (17) and (18) agree well with the simulation results.

Fig. 3 shows the calculated and the simulated values of electric field along the channel length at the drain end for the DMDG SOI MOSFET and the simulated values for DG SOI

MOSFET for the same channel length. Because of the discontinuity in the surface potential of the DMDG structure, the peak electric field at the drain is reduced substantially, by approximately 40%, when compared with that of the DG structure that leads to a reduced hot carrier effect. The agreement between the model (20) and simulated results proves the accuracy of the model.

In Fig. 4, the threshold voltage of DMDG structure as a function of channel length is compared with that of the DG MOSFET and the proposed model (29) with $L_1$ fixed at 50 nm. It can be observed clearly that the proposed DMDG structure exhibits a desired threshold voltage "roll-up", while the threshold voltage of the DG structure rolls-down with the decreasing channel lengths for a fixed $L_1$. This is due to the increase in the $L_1/L_2$ ratio for the decreasing channel lengths and the portion of the larger work function gate is increased as the channel length reduces. This unique feature of the DMDG structure is an added advantage when the device dimensions are continuously shrinking. With the decreasing channel lengths, it is very difficult to obtain precise channel lengths across the wafer. However, a threshold voltage variation from device to device is least desirable. DMDG structure exhibits a threshold voltage which is almost constant with decreasing channel lengths. From the results it is clearly seen that the calculated values of the analytical model tracks the simulated values very well. Fig. 5 shows the DIBL variation along the channel for both the DMDG and the DG SOI MOSFETs for $L_1 = L_2$. The simulated DIBL results are calculated as the difference between the linear threshold voltage ($V_{th,lin}$) and the saturation threshold voltage ($V_{th,sat}$). The parameters, $t_f$, $t_b$ and $t_{si}$ used here are 2nm, 3nm and 20nm respectively and have been chosen to get better characteristics. The linear threshold voltage, is based on the maximum transconductance method at $V_{DS} = 0.05V$. The saturation threshold voltage is based on a modified constant-current method at $V_{DS} = 1V$ where the critical current is defined as the drain current when $V_{GS} = V_{th,lin}$ [30]. Again it can be observed clearly that the DIBL increase in the DMDG structure is far less when compared with the DG MOSFET with the decreasing channel lengths.

The drain current characteristics of both the DMDG and the DG MOSFETs are shown in Fig. 6 for a channel length of 100 nm ($L_1 = L_2 = 50$ nm). In the case of the DMDG structure, the results obtained from the model are also shown. Fig. 6 demonstrates that the DMDG structure exhibits an improved transconductance, reduced drain conductance and an increase in the drain breakdown voltage. This enhancement in the performance is because of the step function of the surface potential profile along the channel, which reduces the DIBL and the peak electric field at the drain end. In the drain current analytical model, various short channel effects such as the channel length modulation, DIBL, velocity overshoot have been considered along with the breakdown mechanisms involved: the parasitic BJT effects and the impact ionization. The enhanced performance is indeed shown in Figs. 7 and 8, where the transconductance ($g_m$) and the drain conductance ($g_d$) for both the structures is plotted with different channel lengths for $L_1 = L_2$. The value of $g_m$ is extracted from the slope of $I_D$-$V_{GS}$ between $V_{GS} = 1$V and 1.5V at $V_{DS} = 0.75$V while $g_d$ is extracted from the slope of $I_D$-$V_{DS}$ between $V_{DS} = 0.5$V and 0.75V at $V_{GS} = 1.5$V for both simulation and model predicted values. Fig. 9 shows the voltage gain of the DMDG and the DG SOI MOSFETs as a function of the channel length for $L_1 = L_2$. Because of an increase in the transconductance and a decrease in the drain conductance, the voltage gain ($g_m/g_d$) of the DMDG structure is much higher when compared with that of the DG structure.

## 6. Conclusions

The concept of Dual-Material-Gate has been applied to the Double Gate SOI MOSFET structure and the features exhibited by the resulting new structure, Dual Material Double Gate structure, have been examined for the first time by developing an analytical model. The results obtained from the model agree well with the MEDICI simulation results. We have demonstrated that the DMDG structure leads to reduced short channel effects as the surface potential profile shows a step at the interface of the two materials of the front gate, which reduces drain conductance and DIBL. Moreover, the peak electric field at the drain end is reduced, minimizing the hot carrier effect. The threshold voltage shows a roll-up with reducing channel lengths. In addition, we have also shown that the DMDG SOI MOSFET offers higher transconductance and improved drain breakdown voltage. All these features should make the proposed DMDG SOI MOSFET a prime candidate for future CMOS ULSI chips. Because of the asymmetric nature of the DMDG structure, it may pose few challenges while integrating with the present CMOS technology. But Zhou [25] suggested two fabrication procedures requiring only one additional mask step with which a dual material gate can be obtained. As the CMOS processing technology is maturing and already into the sub-100 nm [31] regime, fabricating a 50 nm feature gate length should not hinder the possibility of achieving the potential benefits and excellent immunity against SCE's that the DMDG SOI MOSFET promises.

**Figure Captions**

Figure 1  Cross-sectional view of (a) DG-SOI MOSFET (b) DMDG-SOI MOSFET

Figure 2.  Surface potential profiles of DMDG and DG-SOI MOSFETs for a channel length $L = 0.1 \mu m$ ($L_1 = L_2 = 0.05 \mu m$).

Figure 3.  Electric-field variation at the drain end along the channel at the Si-SiO$_2$ interface.of DMDG and DG SOI MOSFETs for a channel length $L = 0.1 \mu m$ ($L_1 = L_2 = 0.05 \mu m$).

Figure 4.  Threshold voltage of DMDG and DG SOI MOSFETs is plotted for different channel lengths ($L_1$ fixed at $0.05 \mu m$.).

Figure 5.  DIBL of DMDG and DG SOI MOSFETs is plotted for different channel lengths, $L = L_1 + L_2$ where $L_1 = L_2$. The parameters used are $t_{ox} = 2nm$ $t_b = 3nm$, $t_{si} = 20nm$.

Figure 6.  $I_D$-$V_{DS}$ characteristics of the DMDG and DG-SOI MOSFETs for a channel length $L = 0.1 \mu m$

Figure 7.  Variation of $g_m$ with different channel lengths, ($L_1 = L_2$) for DMDG and DG SOI MOSFETs.

Figure 8.  Variation of $g_d$ with different channel lengths, ($L_1 = L_2$) for DMDG and DG SOI MOSFETs.

Figure 9.  Variation of voltage gain with different channel lengths, ($L_1 = L_2$) for DMDG and DG SOI MOSFETs.

Table 1

| Parameter | Value |
|---|---|
| Front gate oxide, $t_f$ | 2 nm |
| Back gate oxide, $t_b$ | 2 nm |
| Film thickness, $t_{si}$ | 12 nm |
| Body doping, $N_A$ | $10^{15}$ cm$^{-3}$ |
| Source/drain doping, $N_D$ | $5\times10^{19}$ cm$^{-3}$ |
| Length of source/drain regions | 100 nm |
| Distance between source/drain contact and gate | 50 nm |
| Work function p$^+$ poly | 5.25 eV |
| Work function n$^+$ poly | 4.17 eV |

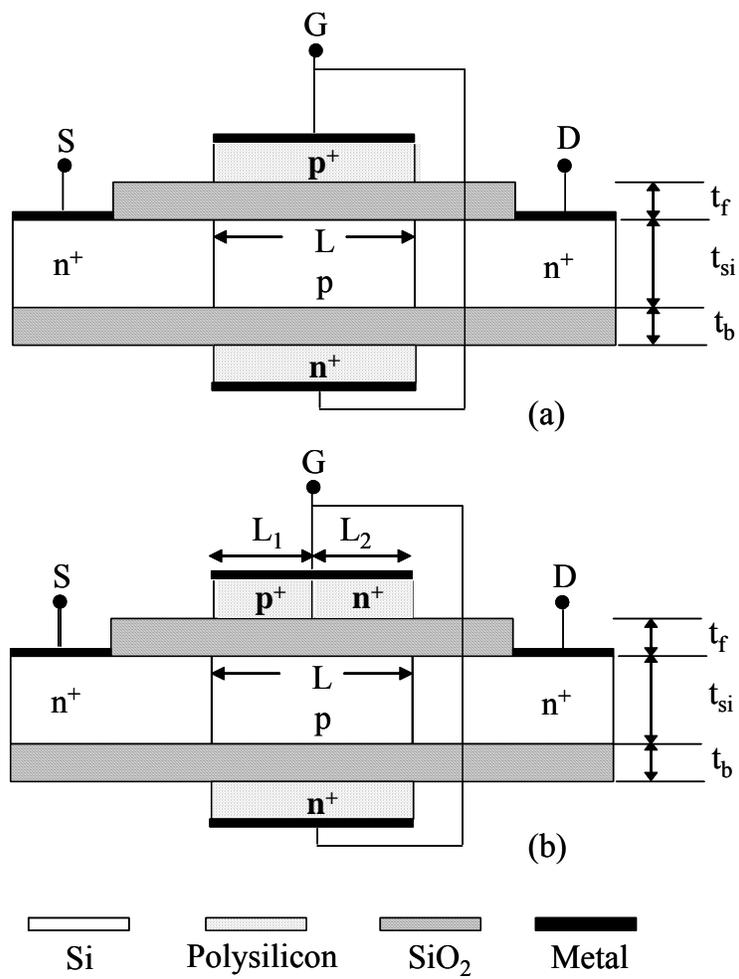

Figure 1

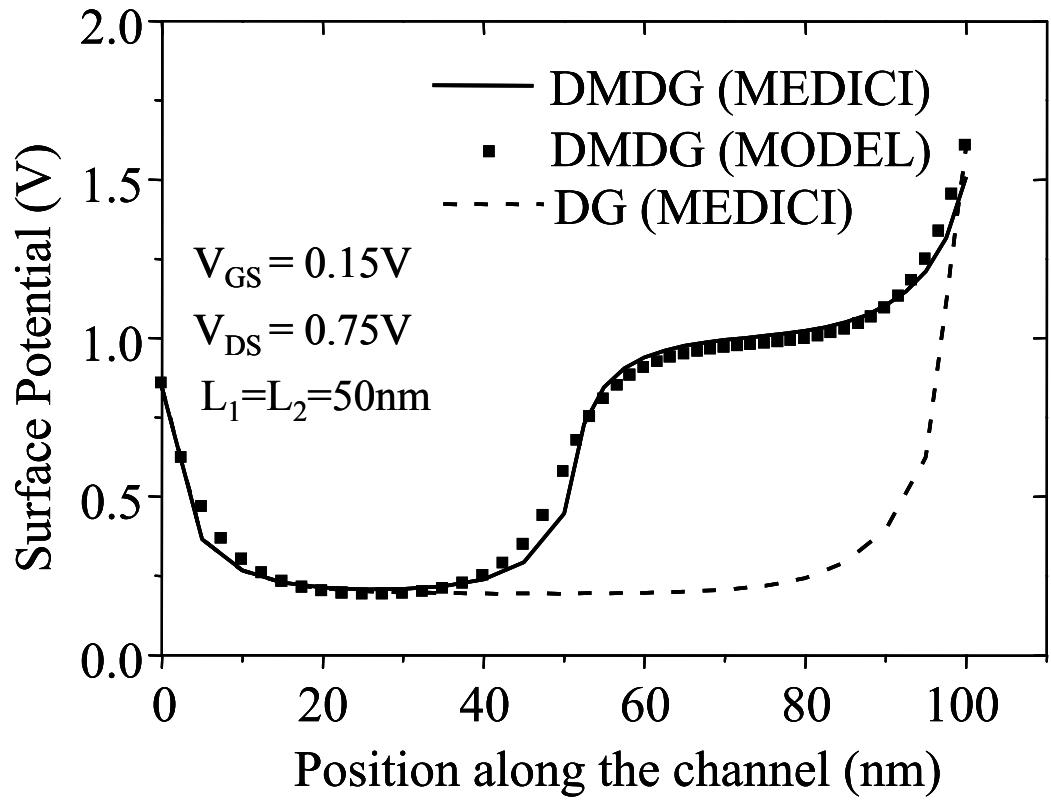

Figure 2

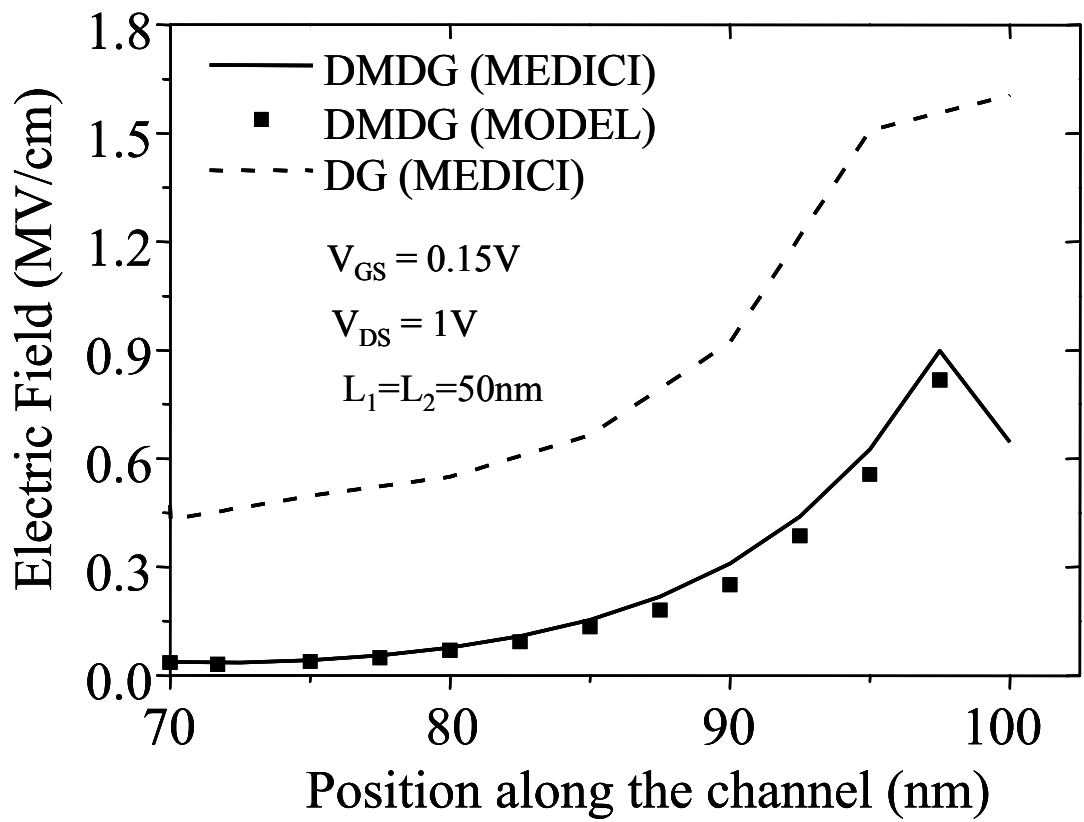

Figure 3

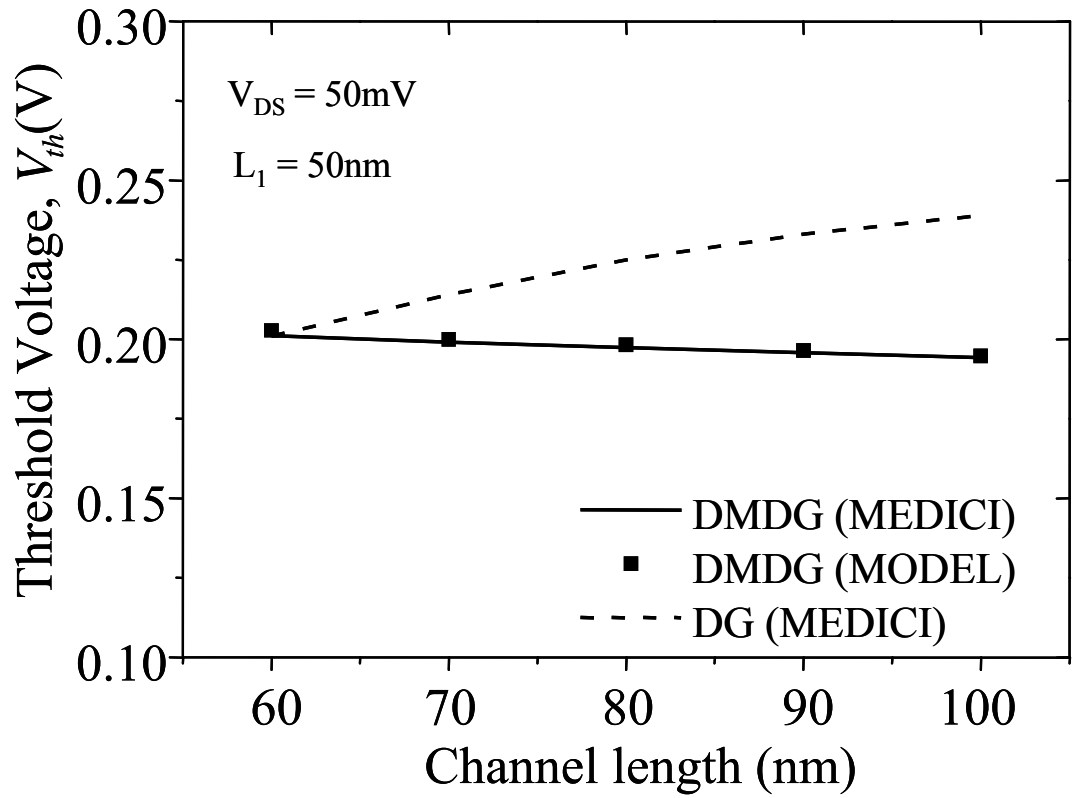

Figure 4

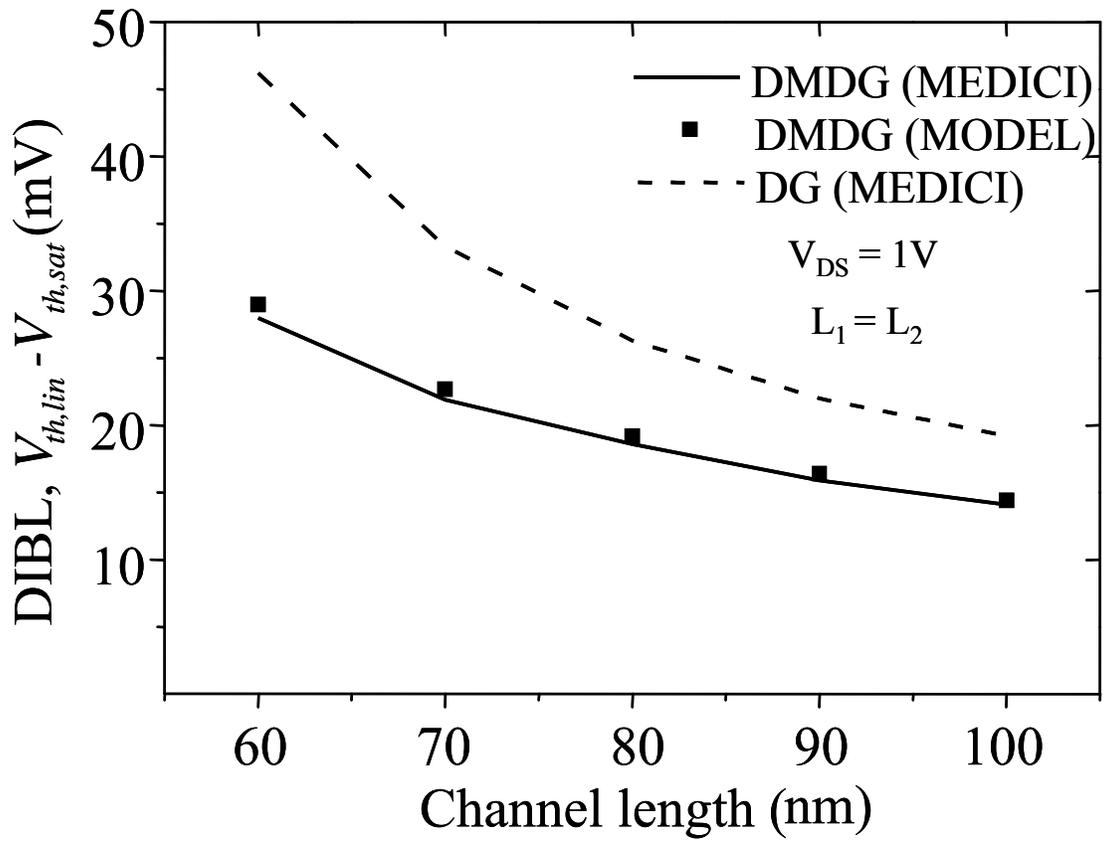

Figure 5

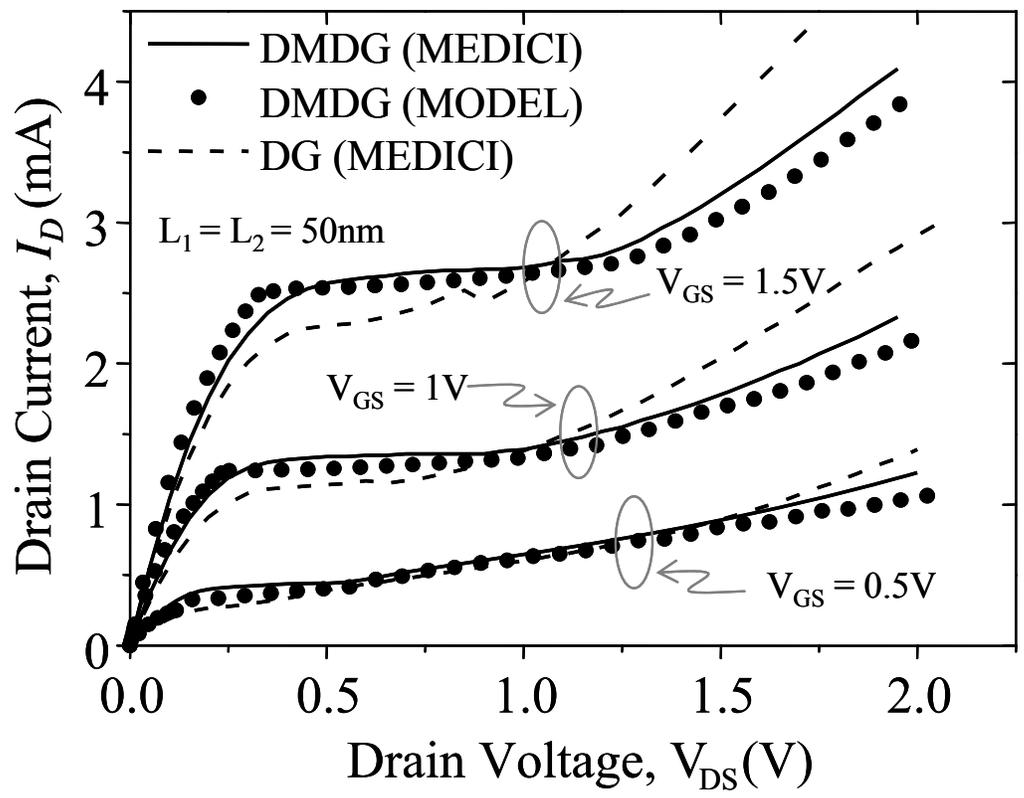

Figure 6

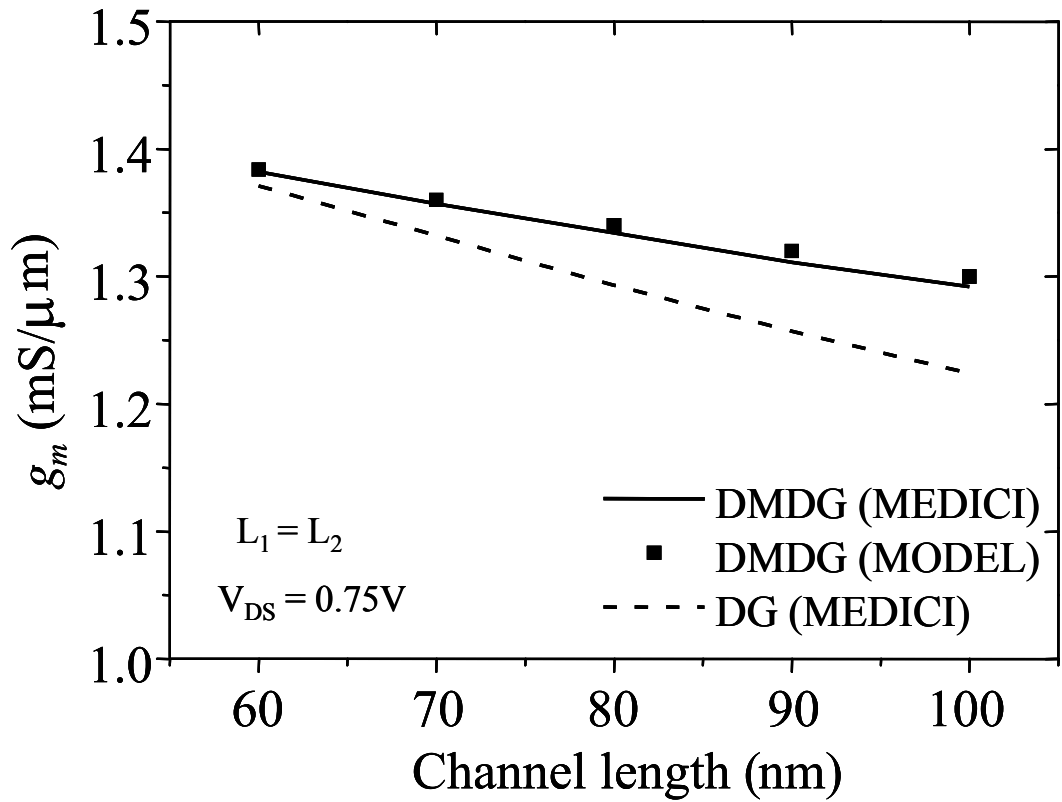

Figure 7

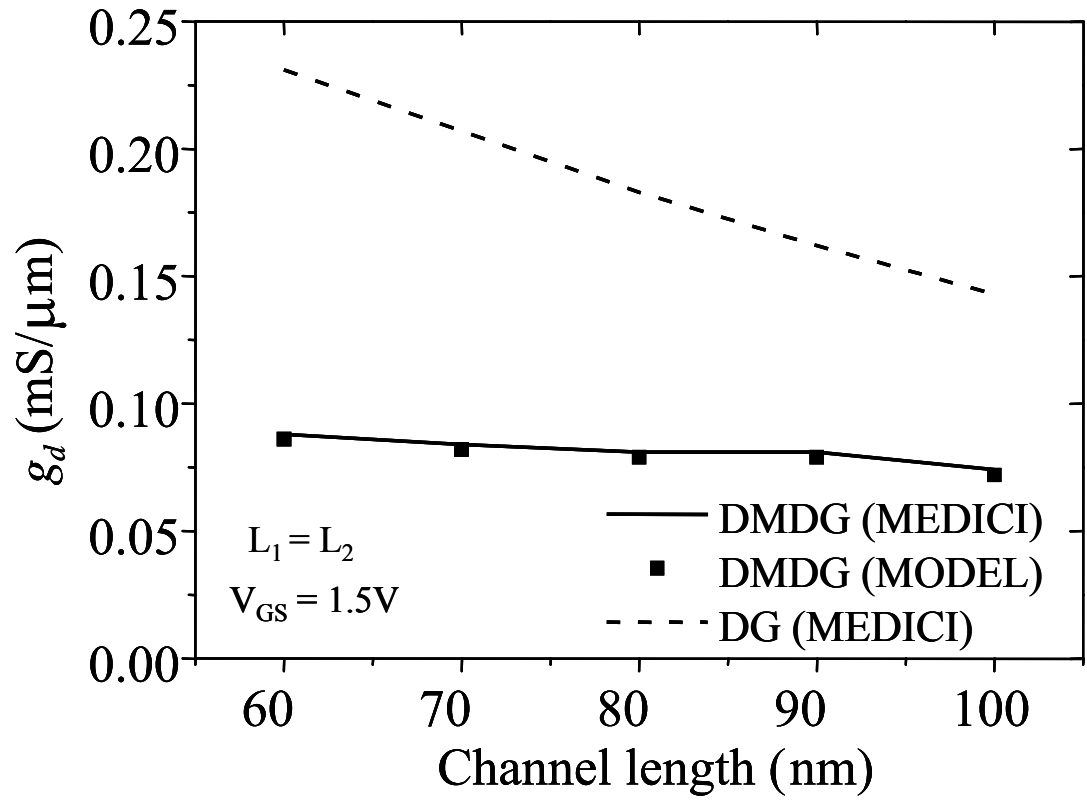

Figure 8

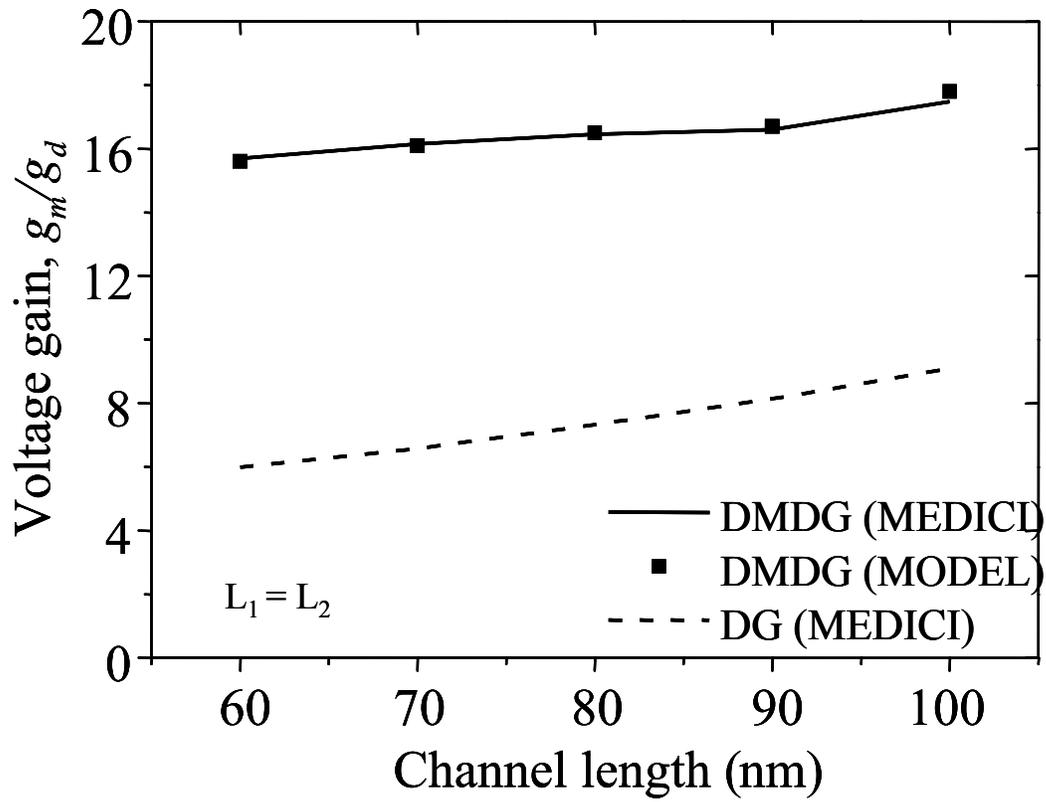

Figure 9